# The long non-coding RNA *HOTAIR* is transcriptionally activated by HOXA9 and is an independent prognostic marker in patients with malignant glioma

Ana Xavier-Magalhães[1,2,*], Céline S. Gonçalves[1,2,*], Anne Fogli[3,4], Tatiana Lourenço[1,2], Marta Pojo[1,2], Bruno Pereira[5], Miguel Rocha[6], Maria Celeste Lopes[7], Inês Crespo[7], Olinda Rebelo[7], Herminio Tão[7], João Lima[8], Ricardo Moreira[8], Afonso A. Pinto[8], Chris Jones[9], Rui M. Reis[1,2,10], Joseph F. Costello[11], Philippe Arnaud[3], Nuno Sousa[1,2] and Bruno M. Costa[1,2]

[1]Life and Health Sciences Research Institute, School of Medicine, University of Minho, Braga, Portugal

[2]ICVS/3B's-PT Government Associate Laboratory, Braga/Guimarães, Braga, Portugal

[3]GReD, Université Clermont Auvergne, CNRS, INSERM, Clermont-Ferrand, France

[4]Biochemistry and Molecular Biology Department, Clermont-Ferrand Hospital, Clermont-Ferrand, France

[5]Biostatistics Department, DRCI, Clermont-Ferrand Hospital, Clermont-Ferrand, France

[6]Centre of Biological Engineering, School of Engineering, University of Minho, Campus de Gualtar, Braga, Portugal

[7]Center for Neuroscience and Cell Biology, University of Coimbra, Coimbra, Portugal

[8]Department of Neurosurgery, Hospital Escala Braga, Braga, Portugal

[9]Divisions of Molecular Pathology and Cancer Therapeutics, The Institute of Cancer Research, Sutton, Surrey, United Kingdom

[10]Molecular Oncology Research Center, Barretos Cancer Hospital, Barretos, São Paulo, Brazil

[11]Department of Neurological Surgery, University of California San Francisco, San Francisco, California, USA

[*]These authors contributed equally to this work

***Correspondence to:*** *Bruno M. Costa, **email:** bfmcosta@med.uminho.pt*

## ABSTRACT

The lncRNA *HOTAIR* has been implicated in several human cancers. Here, we evaluated the molecular alterations and upstream regulatory mechanisms of *HOTAIR* in glioma, the most common primary brain tumors, and its clinical relevance. *HOTAIR* gene expression, methylation, copy-number and prognostic value were investigated in human gliomas integrating data from online datasets and our cohorts. High levels of *HOTAIR* were associated with higher grades of glioma, particularly IDH wild-type cases. Mechanistically, *HOTAIR* was overexpressed in a gene dosage-independent manner, while DNA methylation levels of particular CpGs in *HOTAIR* locus were associated with *HOTAIR* expression levels in GBM clinical specimens and cell lines. Concordantly, the demethylating agent 5-Aza-2'-deoxycytidine affected *HOTAIR* transcriptional levels in a cell line-dependent manner. Importantly, *HOTAIR* was frequently co-expressed with *HOXA9* in high-grade gliomas from TCGA, Oncomine, and our Portuguese and French datasets. Integrated *in silico* analyses, chromatin immunoprecipitation, and qPCR data showed that HOXA9 binds directly to the promoter of *HOTAIR*. Clinically, GBM patients with high *HOTAIR* expression had a significantly reduced overall survival, independently of other prognostic variables. In summary, this work reveals *HOXA9* as a novel direct regulator of *HOTAIR*, and establishes *HOTAIR* as an independent prognostic marker, providing new therapeutic opportunities to treat this highly aggressive cancer.

## INTRODUCTION

Gliomas encompass the majority of primary central nervous system tumors, and have been historically classified based on their histopathological features, according to four grades of malignancy (World Health Organization – WHO – grade I-IV). Glioblastoma (GBM; WHO grade IV) represents the most common and malignant primary adult brain tumor [1], with a median patient overall survival (OS) of approximately 15 months [2]. The remarkable tumor heterogeneity, invasive phenotype, and resistance to radio- and chemotherapy are main determinants of this poor prognosis [3–5]. Additionally, several molecular alterations (epigenetic, genetic, transcriptional and proteomic) contribute to GBM tumorigenesis and may represent putative biomarkers of GBM prognosis and aggressiveness [6].

Recent studies have explored the complexity of the transcriptome and reported the existence of a large number of transcripts that do not code for a protein, which nonetheless have the critical ability to regulate gene expression [7]. Among others, these transcripts include long non-coding RNAs (lncRNA) [8, 9] that are defined as spliced, polyadenylated and >200 nucleotides long RNAs that modify the genome in a highly tissue-specific manner [9, 10]. LncRNA have also emerged as new players in cancer development and progression, with recent studies highlighting the importance of both tumor suppressor and oncogenic lncRNAs in the deregulation of cellular pathways and gene expression programs [11]. The *HOX Transcript Antisense Intergenic RNA* (*HOTAIR*) is a *trans*-acting lncRNA, involved in the genome-wide reprogramming of the chromatin [10]. *HOTAIR* is transcribed at the *HOXC* locus in chromosome 12 (between *HOXC11* and *HOXC12*), and its 5′-domain interacts with the Polycomb Repressor Complex 2 (PRC2), and the 3′-domain interacts with the lysine specific demethylase 1/REST corepressor 1/RE1-silencing transcription factor (LSD1/CoREST/REST) complex. Thus, *HOTAIR* can act as modular scaffold, with different binding sites for harboring demethylase (LSD1, that mediates the enzymatic demethylation of histone H3 dimethyl lysine 4, H3K4me2) and methylase (PRC2 and the histone H3 lysine 27 – H3K27 – methylase EZH2) complexes, resulting in the transcriptional inhibition of the *HOXD* locus on chromosome 2 by epigenetic control of the chromatin structure [10, 12].

*HOTAIR* has also been shown to have functional roles in a variety of human cancers (reviewed in [13–15]), being associated with patients' decreased OS, increased metastatic potential, tumor recurrence and chemotherapy resistance. In glioma, *HOTAIR* was reported to be involved in proliferation, invasion, cell cycle and colony formation ability, *in vivo* tumor growth, and GBM patients' OS [16–20]. However, the molecular mechanisms underlying its aberrant activation in glioma remain elusive. In this context, we aimed to investigate the integrated molecular status (gene copy number, expression, and DNA methylation levels) of *HOTAIR* in glioma samples, unravel the underlying mechanisms regulating *HOTAIR* expression in these tumors, and assess its clinical significance in independent large cohorts of GBM.

## RESULTS

### *HOTAIR* expression is associated with high-grade gliomas

In a first approach, we analyzed the expression levels of *HOTAIR* in grades II and III gliomas, and in grade IV GBM patients using gene expression microarray data and RNA-sequencing (RNA-seq) data from The Cancer Genome Atlas (TCGA). Globally, *HOTAIR* was highly expressed in GBM, particularly in IDH-wt GBM, as compared to lower-grade gliomas and normal brain (Figure 1A and Supplementary Figure 1A). Of note, in the RNA-seq data, IDH-wt grade III gliomas presented higher *HOTAIR* expression than the remaining grade II and III gliomas entities (Supplementary Figure 1A). Concordantly, four additional datasets from the Oncomine database (Murat [21], Freije [22], Sun [23], and Phillips [24]) and two independent Portuguese and French glioma series, confirmed that *HOTAIR* expression is far more frequent in grades III and IV than in grade II gliomas and normal brain (Supplementary Table 1 and Supplementary Figure 1B). In all tested glioma datasets, *HOTAIR* expression was not significantly associated with any other clinicopathological feature (patient age, sex, Karnofsky performance status [KPS], and exposure to radiotherapy or chemotherapy, except for associations with patient age in Phillips [24], Repository of Molecular Brain Neoplasia Data (REMBRANDT) and the French datasets, and associations with IDH mutation status in the French datatset; data not shown). Moreover, *HOTAIR* expression was not associated with any particular GBM molecular subtype as defined by Wang *et al.* [25] (Supplementary Figure 1C and 1D; $p > 0.9999$ for microarray and RNA-seq data). Together, these data suggest that *HOTAIR* is highly expressed in high-grade gliomas, being particularly frequent in IDH-wt cases.

**DNA methylation regulates *HOTAIR* levels in glioma**

In order to understand the molecular mechanisms underlying *HOTAIR* increased expression in GBMs, we analyzed gene copy number aberrations (Figure 1B) and DNA methylation levels in the *HOTAIR* locus (Figure 1C). In 250 IDH-wt GBMs analyzed for gene copy number, *HOTAIR* amplification and deletion (either focal or chromosomal) were found in 0.8% (2/250) and in 3.2% of the samples (8/250; Figure 1B), respectively, while no amplifications or deletions were found in 20 IDH-mut GBMs (Figure 1B). While *HOTAIR* levels were high in 3 of the 5 *HOTAIR*-amplified tumors, it was also present in 1 tumor with gene deletion and in 116 without copy number aberrations (Figure 1B). These data show *HOTAIR* copy number aberrations are rare in GBM, and unlikely associated with its increased expression levels ($p = 0.182$). Analysis of 56 methylation probes spanning from *HOTAIR*'s most proximal genes (*HOXC11* and (*HOXC12*) revealed a wide spectrum of methylation levels across different GBM samples (Figure 1C). Of note, some probes consistently presented similar methylation levels across all samples (e.g., several probes within the *HOTAIR* intragenic region consistently showed low methylation levels), suggesting those regions are not prone to *de novo* methylation in GBM. In a subset of 70 GBMs (66 IDH-wt and 4 IDH-mut) for which both gene expression and methylation data were available, we found the methylation levels of 3 probes located in the intragenic region of *HOTAIR* (cg00079219, cg18824990 and cg24895871) to be significantly inversely correlated with *HOTAIR* expression levels ($r = -0.36$, $p = 0.002$; $r = -0.43$, $p = 0.0002$; and $r = -0.35$, $p = 0.003$, respectively; Figure 1D). Of note, performing the same analysis exclusively for the IDH-wt GBMs showed that the methylation levels of the probes cg00079219 and cg18824990 were inversely correlated with *HOTAIR* expression ($r = -0.39$, $p = 0.0002$; $r = -0.39$, $p = 0.0014$), while methylation levels at the probe cg24895871 lost their association with expression ($r = -0.26$, $p = 0.167$). We further corroborated the relevance of these associations in IDH-wt GBM patients in the French dataset as the 3 intragenic methylation probes were also found to inversely correlate with *HOTAIR* expression (cg00079219 $r = -0.53$, $p = 0.016$; cg18824990 $r = -0.523$, $p = 0.017$; cg24895871 $r = -0.521$, $p = 0.018$), and in the combination of IDH-wt glioma grade III and IDH-wt GBM patients (cg00079219 $r = -0.43$, $p = 0.004$; cg18824990 $r = -0.453$, $p = 0.002$; cg24895871 $r = -0.431$, $p = 0.004$).

To further explore the relevance of DNA methylation in *HOTAIR* regulation, we treated glioma cell lines with the global DNA demethylating agent 5-Aza-2′-deoxycytidine (5-Aza) followed by methylation-specific PCR (MSP) and quantitative PCR (qPCR) analyses (Figure 1E and 1F). Interestingly, *HOTAIR* intragenic CpGs were found to be mostly methylated in untreated glioma cells. In addition, 5-Aza treatment resulted in DNA demethylation (Figure 1E) and associated increase in *HOTAIR* transcriptional levels in most but not all of the cell lines (Figure 1F), suggesting a cell line-dependent regulation. Together, our data suggest an association between methylation and *HOTAIR*'s expression in GBM patients and glioma cell lines, but other mechanisms may be crucial in reactivating *HOTAIR* in these tumors.

**HOXA9 directly binds the *HOTAIR* promoter in GBM cells**

Previously published transcriptomic data from our group (GEO accession number GSE56517) [26, 27] suggested a possible regulation of *HOTAIR* by the homeoprotein HOXA9, an important protein in the aggressiveness, chemotherapy resistance, and prognosis of GBM [21, 26, 28, 29]. Specifically, GBM cell lines with overexpression or silencing of HOXA9 presented increased or reduced expression of *HOTAIR*, respectively; raising the hypothesis that HOXA9 may directly regulate *HOTAIR* expression. Indeed, an *in silico* analysis using MatInspector (Genomatix) revealed 3 putative binding sites for HOXA9 with high matrix similarities to HOXA9.02 (0.899), PBX-HOXA9.01 (0.849), and MEIS1B-HOXA9.01 (0.837) matrixes (Figure 2A), suggesting that HOXA9 may bind to the promoter region of *HOTAIR* and regulate its transcription. To test this, we studied the expression pattern of *HOTAIR* and *HOXA9* in adult and pediatric glioma-derived cell lines by semi-quantitative reverse transcription-PCR (RT-PCR) and qPCR, and found these genes to be frequently co-expressed in glioma (Figure 2B and 2C). Moreover, U87MG-HOXA9 cells (with exogenous overexpression of HOXA9) presented significantly increased expression of *HOTAIR* as compared to their HOXA9-negative counterpart (U87MG-MSCV; Figure 2B and 2C). We then performed anti-HOXA9 chromatin-immunoprecipitation (ChIP) on U251 GBM cell line (endogenously expressing *HOTAIR* and *HOXA9*), U87MG-HOXA9 (*HOXA9* overexpression models) and their HOXA9-negative counterparts. In both U251 and U87MG-HOXA9, ChIP-qPCR revealed significantly increased HOXA9 occupancy of the promoter region of *HOTAIR* in comparison to their respective controls (Figure 2D and 2E; $p < 0.0002$ for U251; $p = 0.0148$ for U87MG-HOXA9), confirming that HOXA9 directly interacts with the promoter region of *HOTAIR*. Together, these results indicate that HOXA9 is a direct activator of *HOTAIR* expression in GBM cells, thus supporting the strong co-expression of these genes observed in glioma cell lines.

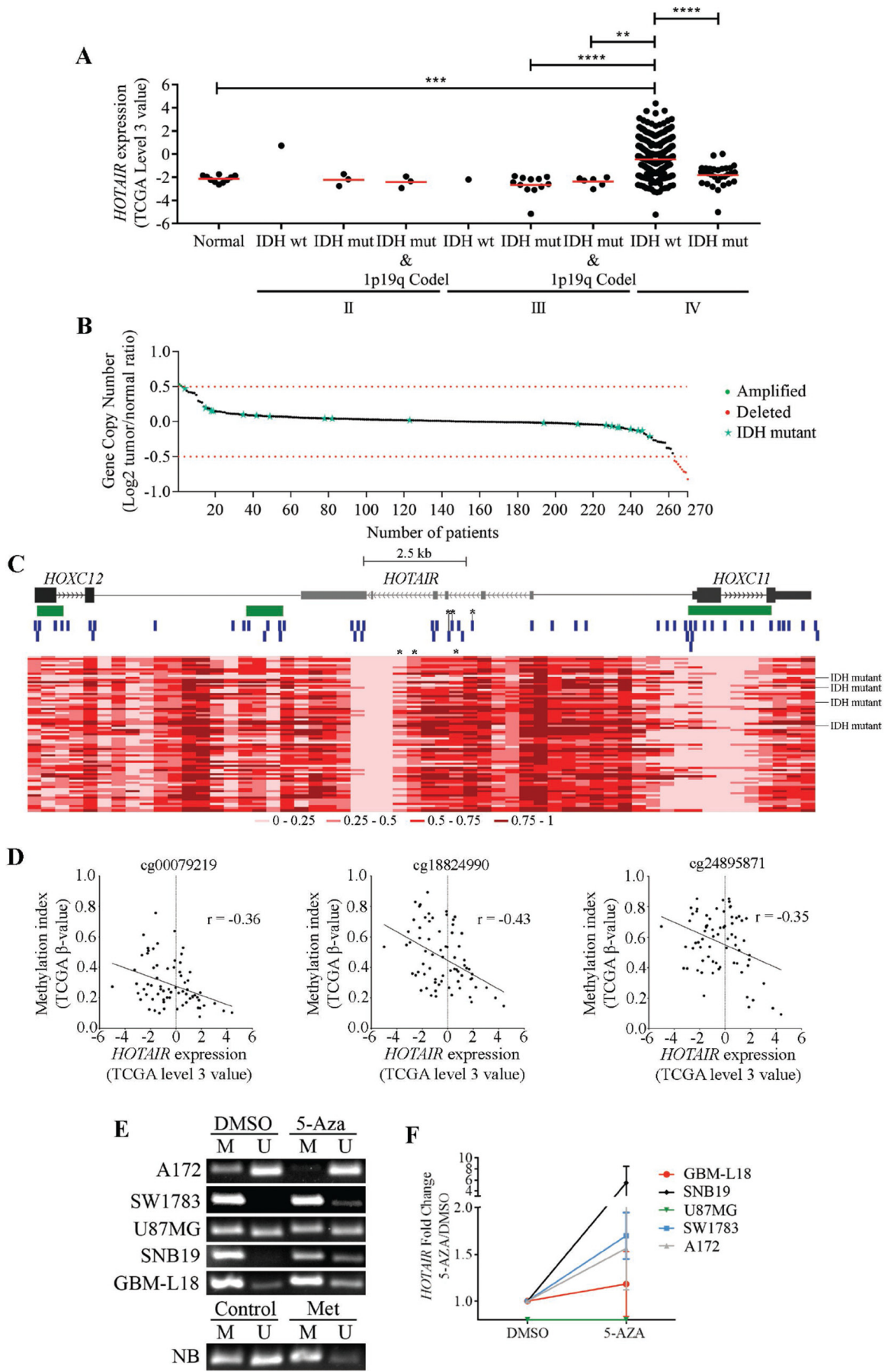

**Figure 1: Molecular characterization of *HOTAIR* in gliomas.** (**A**) Expression levels of *HOTAIR* in 424 gliomas, stratified according to WHO grade, IDH and 1p/19q codeletion statuses (1 IDH-wt, 3 IDH-mut, and 3 IDH-mut and 1p/19q codeleted grade II gliomas; 1 IDH-wt, 12 IDH-mut, and 6 IDH-mut and 1p/19q codeleted grade III gliomas; 368 IDH-wt and 30 IDH-mut glioblastomas (GBM); and 10 unmatched normal brains from the TCGA microarray data). *HOTAIR* is highly expressed (TCGA data "level 3" values ≥ 0) in 34.2% ($n$ = 126) of IDH-wt GBM samples and in 1 IDH-mut GBM (3%) and 1 IDH-wt grade II glioma (100%). (**B**) *HOTAIR* gene copy number status in 270 GBMs (250 IDH-wt and 20 IDH-mut) from TCGA. *HOTAIR* is amplified ($Log_2$ Copy Number Tumor/Normal ≥ 0.5) in 0.8% ($n$ = 2; green dots), and deleted ($Log_2$ Copy Number Tumor/Normal ≤ −0.5) in 3.2% ($n$ = 8; red dots) of IDH-wt GBM samples. Red dashed lines represent the normal copy number interval. (**C**) Heatmap representations of DNA methylation levels (TCGA β-values) of the chromosomal region encompassing *HOTAIR* and the 2 closest genes (*HOXC12* and *HOXC11*) in 74 GBMs (70 IDH-wt and 4 IDH-mut) from TCGA. A total of 56 methylation probes (vertical blue bars) were assessed. CpG islands > 300bp are represented in green. *indicate probes whose methylation indexes are significantly inversely correlated with *HOTAIR* expression levels (probes cg00079219, cg18824990 and cg24895871). The color code (grades of red color corresponding to different methylation indexes) is shown below the heatmap. Each column corresponds to a probe and each row to a patient. (**D**) Correlation graphs between *HOTAIR* expression levels (TCGA "level 3" value) and DNA methylation indexes (TCGA β-values) in 70 GBM samples. Only probes whose methylation values are significantly inversely correlated with *HOTAIR* expression are shown (cg00079219, cg18824990 and cg24895871; marked with *in C). (**E–F**) Glioma cell lines were treated with 5 μM 5-Aza for 72 hours, upon which promoter methylation status (E) and *HOTAIR* expression levels (F) were evaluated. 5-Aza treatment promoted *HOTAIR* promoter demethylation (E) that associated with its increased expression in a cell line-dependent manner (F). qPCR levels were normalized to the expression of *HPRT1* and are presented as fold-changes; methylation-specific PCR was controlled by blood DNA (NB) untreated (Control) or *in vitro* methylated (Met). No detectable *HOTAIR* expression was found for U87 (untreated or 5-Aza-treated). The results are representative of at least 2 replicates (mean ± SD). *$p$ < 0.05; NB - DNA from normal blood.

## Co-expression of *HOTAIR* and *HOXA9* is a frequent event in clinical specimens of glioma

To verify if the frequent co-expression of *HOTAIR* and *HOXA9* in glioma cell lines is present in glioma clinical specimens, we evaluated their expression levels and potential correlations in the Portuguese, French, and TCGA datasets. We found that *HOTAIR* and *HOXA9* frequently have a coherent transcriptional activation status in grade IV glioma (Portuguese dataset Pearson's $r$ = 0.642, $p$ < 0.0001; IDH-wt GBM in French dataset Pearson's $r$ = 0.649, $p$ = 0.002; IDH-wt GBM in TCGA microarray dataset Pearson's $r$ = 0.494, $p$ < 0.0001; IDH-wt GBM in TCGA RNA-seq dataset Pearson's $r$ = 0.525, $p$ < 0.0001; Figure 3A–3C, Supplementary Figure 2H and Table 1). Similarly, significant correlations between *HOTAIR* and *HOXA9* expression were also observed in specific combinations of glioma grades (Portuguese dataset WHO grades II+III, III, III+IV and II+III+IV; French dataset IDH-wt WHO grade III and grade IV; Table 1 and Supplementary Figure 2A–2G). While *HOTAIR* and *HOXA9* co-expression did not reach statistical significance for grade II gliomas ($n$ = 6; Supplementary Figure 2A), this is likely due to the small sample size, and the same trend for co-expression was observed; indeed, all of these tumors were either double-positive or double-negative (Table 1).

Further analysis in 4 independent datasets showed a statistically significant correlation between the expression of *HOTAIR* and *HOXA9* in all glioma sets (Table 2). Furthermore, in IDH-wt ($n$ = 55) and IDH-mut ($n$ = 137) grades II and III glioma with available RNA-seq data from TCGA, a statistically significant correlation between the expression of *HOTAIR* and *HOXA9* was also found (IDH-wt Spearman's $r$ = 0.65, $p$ < 0.0001, and IDH-mut $r$ = 0.21, $p$ = 0.01; data not shown), while in 1p19q codeleted and IDH-mut grades II and III glioma no correlation was found ($n$ = 85, $r$ < 0.01, $p$ = 0.999; data not shown).

Interestingly, no significant correlations between *HOTAIR* and *HOXA9* were found for other cancer types, including lung, leukemia, colorectal, and breast cancers (Table 2). Taken together, our results strongly suggest that *HOTAIR* and *HOXA9* are concomitantly co-expressed in human gliomas, but not in other cancer types.

## High levels of *HOTAIR* expression associate with shorter survival in patients with malignant glioma

To validate the prognostic value of *HOTAIR* expression reported in GBM patients from the Chinese population [16], we firstly analyzed two independent datasets of GBM patients from TCGA and REMBRANDT. In 554 GBM patients from TCGA with available survival data, a statistically significant decrease in OS was observed in patients with high *HOTAIR* expression ($n$ = 177, median OS = 383 days) as compared to patients whose GBMs presented low *HOTAIR* levels ($n$ = 377, median OS = 447 days; $p$ = 0.026 Log-rank test; Figure 4A). Similarly, associations were found in the subset of 387 IDH-wt GBMs (*HOTAIR*-high $n$ = 126, median OS = 370 days and *HOTAIR*-low $n$ = 241, median OS = 447 days; $p$ = 0.032 Log-rank test; Figure 4B and Supplementary Table 2). As expected, classic prognostic factors including patient age at diagnosis, KPS, and use of radio-chemotherapy, were significantly associated with OS ($p$ < 0.05, Log-rank test, Supplementary Table 2). Importantly, *HOTAIR* expression was significantly associated with shorter OS of IDH-wt GBM patients in a multivariable Cox model ($p$ = 0.036), independently of other prognostic factors (age, gender, KPS, use of chemoradiotherapy, and treatment with additional chemoradiotherapy; Supplementary Table 2). Consistently, in an independent dataset of 67 GBM patients from the REMBRANDT dataset, high *HOTAIR* expression was present in 72% of patients ($n$ = 48) who

www.oncotarget.com                                                                 15744                                                                     Oncotarget

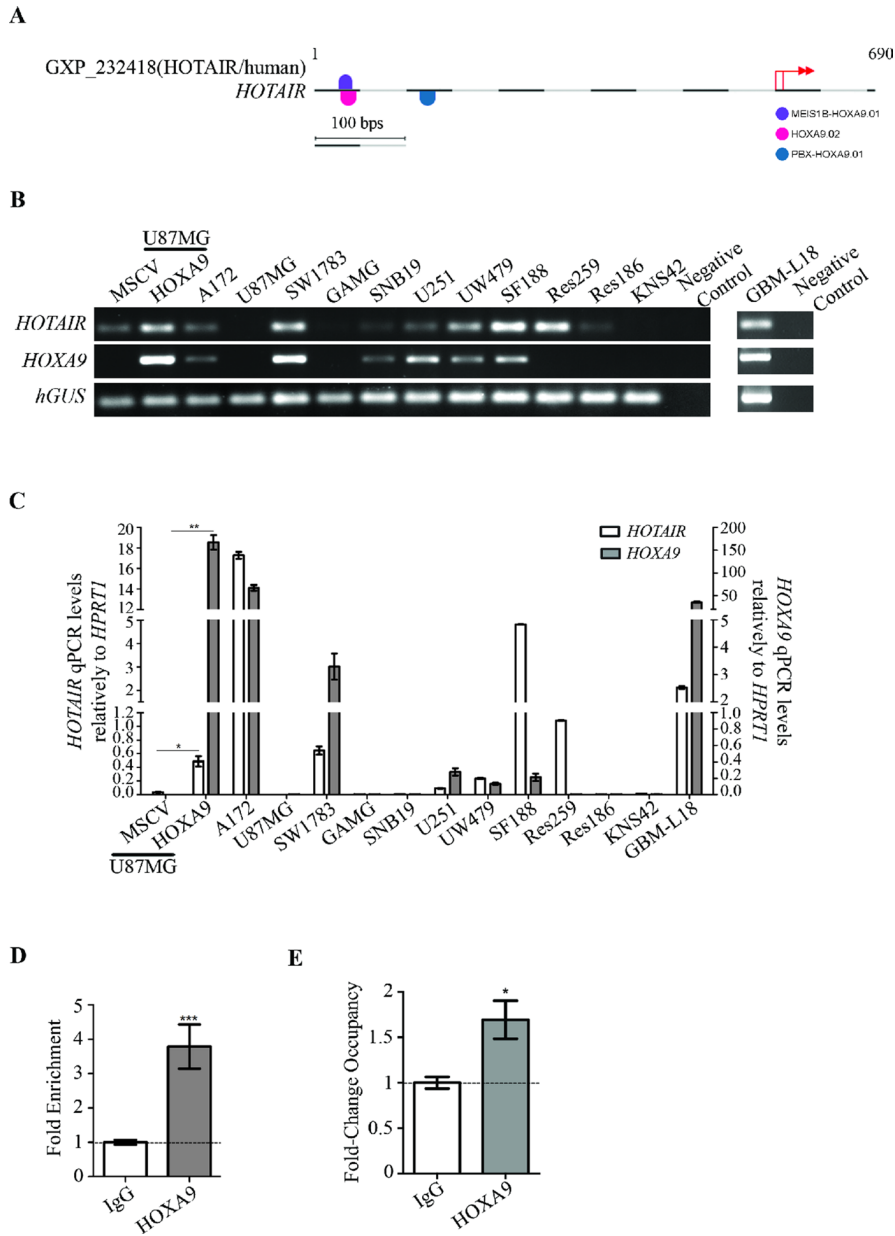

**Figure 2: HOXA9 transcriptionally activates *HOTAIR* via direct interaction with its promoter region.** (**A**) MatInspector representation of specific binding sites for HOXA9 matrix in *HOTAIR* promoter region. Each matrix match is represented by a half round symbol and each color symbolizes a matrix family. Pink, blue and violet colors represent the matrixes HOXA9.02 (matrix sim = 0.899, sequence: ctgtgac*aTAAA*attgg and family ABDB), PBX-HOXA9.01 (matrix sim = 0.849, sequence: gag*gTGGTttat*gagct and family HOXC) and MEIS1B-HOXA9.01 (matrix sim = 0.837, sequence: *TGCCa*at*ttt*atgtc and family HOXH), respectively. Basepairs in italic appear in a position with a high conservation profile in the matrix (ci-value > 60). Basepairs in capital letters represent the core sequence used by the program. Matches represented on the top of the sequence line were found on the positive strand, while below the sequence line reside matches found on the negative strand. The red arrows represent *HOTAIR* putative transcription start sites (TSS). (**B–C**) *HOTAIR* and *HOXA9* expression were evaluated by reverse-transcriptase PCR (B) and quantitative PCR (C) in a panel of adult and pediatric glioma-derived cell lines, and in the GBM-L18 primary GBM-derived cell culture. GBM cell line U87MG-MSCV does not present detectable levels of endogenous *HOTAIR* expression, which were significantly increased upon retrovirally-mediated *HOXA9* overexpression (U87MG-HOXA9). White and grey bars represent *HOTAIR* and *HOXA9* expressions, respectively. qPCR levels were normalized to the expression of *HPRT1*. The results are representative of triplicates (mean ± SD). $^{*}p$ = 0.029; $^{**}p$ = 0.0088. (**D–E**) The putative binding of HOXA9 protein to the promoter region of *HOTAIR* was assessed by chromatin immunoprecipitation (ChIP) analysis followed by quantitative PCR in U251 cells (D), and U87MG-HOXA9 and their *HOXA9*-negative counterparts (E). IgG was used as negative control for the ChIP. Chromatin immunoprecipitated with an anti-HOXA9 antibody shows direct binding of HOXA9 to the *HOTAIR* promoter. Relative enrichment is normalized to input DNA (not subjected to immunoprecipitation) and to the IgG background signal (D), whereas fold change occupancy is normalized to input, IgG and *HOXA9*-negative cells (E), from three independent experiments (mean ± SD). (D) $^{***}p$ = 0.0002; (E) $^{*}p$ = 0.0148.

had a significantly shorter OS (median OS = 15.8 months) than those with low expression (median OS = 37.4 months), both in univariable and multivariable analyses (*p* = 0.005 Log-rank test; *p* = 0.034 Cox regression, Figure 4C and Supplementary Table 3). Since *HOXA9* expression is a known prognostic factor in GBM patients [26, 28, 29], we investigated in the large TCGA dataset whether *HOTAIR* was still prognostically valuable in the subset of *HOXA9*-negative IDH-wt GBM patients. Indeed, high expression of *HOTAIR* was able to identify a subgroup of *HOXA9*-negative IDH-wt GBM patients that presented a significantly shorter OS (median OS = 375 days) than those with low *HOTAIR* levels (median OS = 463 days; Supplementary Figure 3; Log-rank *p* = 0.037).

Additionally, we tested the prognostic value of *HOTAIR* expression among 28 glioma grade III patients from the French dataset with available survival data. Similarly to GBM, we found for the first time an association between high *HOTAIR* expression levels and shorter OS in grade III glioma patients with high *HOTAIR* expression (*p* = 0.002, Log-rank test; Supplementary Table 4). Multivariable Cox analysis further showed that high *HOTAIR* expression is a poor prognostic factor (*p* = 0.022) independently of other known prognostic factors such as age at diagnosis, surgery (partial or complete resection), KPS, treatment, and IDH and 1p19q status. The median survival is 1.4 years [0.7–2.5] in the group of 28 grades III. Survival medians are respectively of 2 years in the group of patients with *HOTAIR*-lower expression (*n* = 16, median *HOTAIR* expression = 0.0002 [2.96e-6 – 0.0053]) and of 0.7 years in the group of patients with *HOTAIR*-higher expression (*n* = 12, median *HOTAIR* expression = 0.0387 [6.4e-6 – 0.0332]).

Finally, we found that high *HOTAIR* expression is also associated with shorter overall survival in grade II

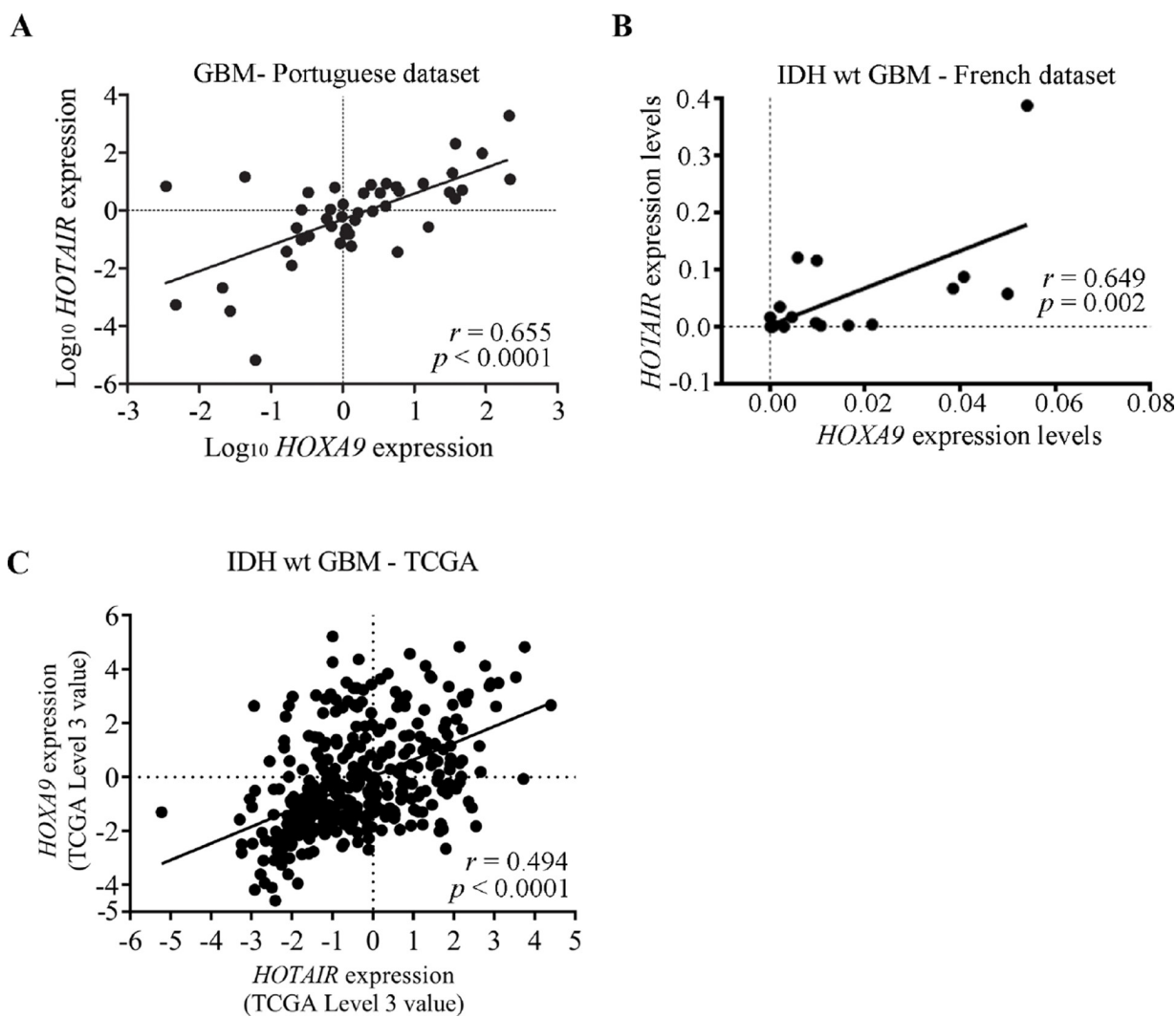

**Figure 3: Expression levels of *HOTAIR* and *HOXA9* are significantly correlated in GBM clinical specimens.** (**A–C**) Correlation graphs of *HOTAIR* and *HOXA9* expression levels in GBMs from the Portuguese dataset (A), and IDH-wt GBMs from the French dataset (B), and IDH-wt GBMs the TCGA dataset (microarray data) (C), showing statistically significant correlations between their expressions in all datasets. (A) Pearson's *r* = 0.655, *p* < 0.0001; (B) Pearson's *r* = 0.649; *p* = 0.002; (C) Pearson's *r* = 0.494; *p* < 0.0001.

**Table 1: Analysis of *HOTAIR* and *HOXA9* expression correlations in human glioma samples from the Portuguese dataset**

|  |  | Grade II | | Grade III | | Grade IV | |
|---|---|---|---|---|---|---|---|
|  |  | *HOTAIR* | | *HOTAIR* | | *HOTAIR* | |
|  |  | + | − | + | − | + | − |
| *HOXA9* | + | 1 | 0 | 4 | 4 | 8 | 6 |
|  | − | 0 | 5 | 1 | 5 | 5 | 34 |
|  |  | *r* = 0.774 *p* = 0.07 | | *r* = 0.546 ***p* = 0.030** | | *r* = 0.655 ***p* < 0.0001** | |
|  |  | | *r* = 0.619 ***p* = 0.005** | | | | |
|  |  | | | | *r* = 0.640 ***p* < 0.0001** | | |
|  |  | | | *r* = 0.642 ***p* < 0.0001** | | | |

*Pearson's correlation.

**Table 2: Analysis of *HOTAIR* co-expression with *HOXA9* in glioma datasets and in other cancer types (lung, leukemia, colorectal, and breast cancer datasets from TCGA) available at Oncomine[(1)]**

| Datasets | Correlation *HOTAIR*/*HOXA9* | *p*-value |
|---|---|---|
| **Glioma, NOS** | | |
|    Portuguese dataset | 0.642 | **<0.0001** |
|    French dataset | 0.516 | **<0.001** |
|    TCGA, microarray data | 0.545 | **<0.0001** |
|    TCGA, RNA-seq data | 0.675 | **<0.0001** |
|    Oncomine: | | |
|       Freije *et al*. (Freije, *et al*. 2004) | 0.372 | **<0.001** |
|       Murat (Murat, *et al*. 2008) | 0.293 | **0.008** |
|       Sun (Sun, *et al*. 2006) | 0.487 | **<0.0001** |
|       Phillips (Phillips, *et al*. 2006) | 0.309 | **0.001** |
| **Other cancers from Oncomine:** | | |
|    TCGA Lung | 0.046 | 0.555 |
|    TCGA Leukemia | −0.019 | 0.791 |
|    TCGA Colorectal | 0.012 | 0.854 |
|    TCGA Breast | −0.051 | 0.215 |

[(1)]Microarray data from Oncomine dataset, microarray or RNA-seq data from TCGA dataset, quantitative PCR data from the Portuguese dataset (relative gene expression normalized to *hGUS* or *TBP* levels), and microfluidic-based qPCR analysis (Fluidigm) data from the French dataset were used to evaluate the Pearson's correlation.
NOS – not otherwise specified.

($n$ = 226) and grade III ($n$ = 240) glioma patients in the RNA-seq data from the larger TCGA dataset (Supplementary Tables 5 and 6; $p$ = 0.020 and $p$ < 0.0001, respectively; Log-rank test), and independently of patient age, gender, and molecular subtype (IDH and 1p/19q status) in the case of grade II gliomas ($p$ = 0.032; Cox regression model), but not for grade III gliomas ($p$ = 0.395; Cox regression model), in which high *HOTAIR* levels were highly correlated with wild-type IDH.

Taken together, our findings establish *HOTAIR* as a clinically-relevant biomarker of prognosis in malignant glioma patients.

## DISCUSSION

An increasing focus has been given to lncRNAs as key regulators of a range of biological functions and as crucial players in the etiology of disease, especially in cancer [8, 9]. The relevance of *HOTAIR* in cancer was first reported by an association between *HOTAIR* expression and increased metastatic potential and diminished survival of breast cancer patients [30]. Later studies evaluated the roles of *HOTAIR* in several tumor types (reviewed in [13–15]), globally highlighting oncogenic roles of *HOTAIR*. Likewise, *HOTAIR* expression was associated with more aggressive disease in glioma patients [16–19]. While some oncogenic functions of *HOTAIR* have been reported in glioma [17, 31, 32], the underlying mechanisms responsible for *HOTAIR* activation in such a devastating disease have remained unknown thus far. Here, we explore the multi-layered complexity of *HOTAIR* activation by integrating molecular (genetic and epigenetic) analyses, *in silico* approaches, and data from glioma patients and cell lines. Additionally, we further investigated the relevance of *HOTAIR* as a prognostic biomarker in GBM, and in malignant WHO grade II and III glioma.

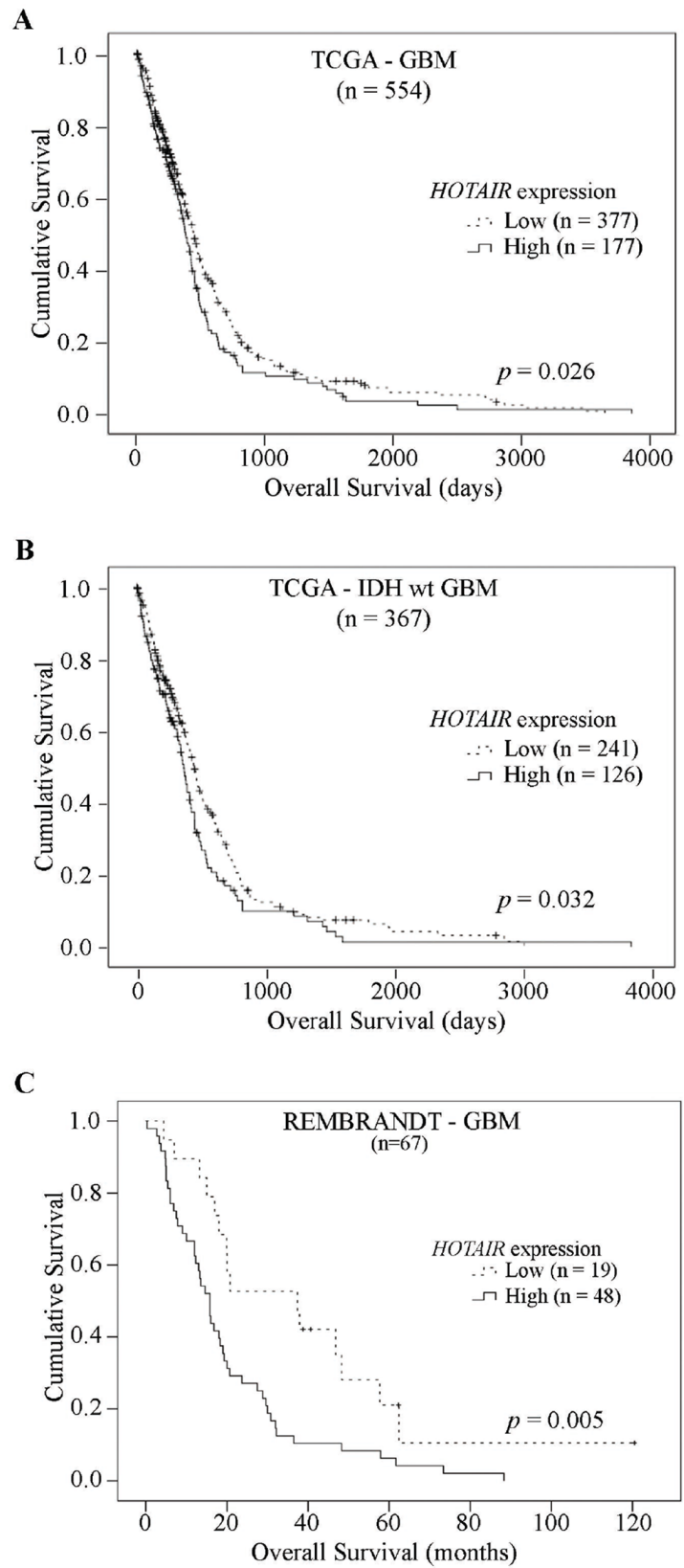

**Figure 4: High levels of *HOTAIR* expression are significantly associated with shorter survival in GBM patients.** Kaplan–Meier overall survival curves of *HOTAIR* prognostic value in (**A**) 554 GBM patients from TCGA, (**B**) 367 IDH-wt GBM patients from TCGA and (**C**) 67 GBM patients from REMBRANDT dataset, showing patients whose tumors present high *HOTAIR* expression have a statistically significant shorter overall survival compared to those with *HOTAIR*-low tumors, in both independent datasets (A, $p = 0.026$; B, $p = 0.032$; C, $p = 0.005$; Log-rank tests).

We show that *HOTAIR* is overexpressed in a subset of high-grade gliomas, particularly in GBMs. Of note, our study includes data from independent patient datasets, and obtained with different complementary techniques, including qPCR, expression microarrays, microfluidic-based qPCR analysis (Fluidigm) and RNA-Seq data. To understand the molecular mechanisms regulating the expression of *HOTAIR*, we evaluated copy number aberrations and DNA methylation levels, and searched for putative transcription factors with binding sites in *HOTAIR*'s promoter region. Copy number alterations in *HOTAIR* locus were rare and did not associate with its expression levels, suggesting this is not a prominent underlying mechanism in GBM. Contrarily, modulation of DNA methylation levels in GBM cell lines by the demethylating agent 5-Aza indicated that intragenic CpGs in *HOTAIR* locus affected its transcriptional levels in a cell line-dependent manner (globally, most cell lines increased *HOTAIR* levels upon demethylation, while U87MG did not, possibly attributable to different kinetics of 5-Aza response and/or lack of expression of all necessary *HOTAIR* transcriptional regulators). Moreover, methylation probes covering CpG islands in *HOTAIR* locus showed significant inverse correlations between *HOTAIR*'s intragenic DNA methylation and expression levels in GBM specimens. In fact, it was previously shown that intragenic DNA methylation is associated with decreased gene expression by altering the chromatin structure and reducing the efficiency of RNA Pol II elongation [33, 34]. Although these results indicate that DNA methylation may regulate *HOTAIR* expression in GBM, this association was not universal, so other transcriptional regulation mechanisms are involved in glioma. Previous observations from our group suggested HOXA9 as a possible regulator of *HOTAIR* in glioma [26]. HOXA9 is an oncogenic transcription factor whose expression was described in a subset of highly aggressive GBMs, and associated with patients' decreased OS [26, 28]. Interestingly, we found *HOTAIR* and *HOXA9* are frequently co-expressed in gliomas, but this was not observed in patients with lung, leukemia, colorectal, and breast cancers, suggesting that this co-expression may be an exclusive event of gliomas. Indeed, our ChIP data indicates a direct regulation of *HOTAIR* by HOXA9, which binds to *HOTAIR* promoter region to activate its transcription. While *HOXA9* and other *HOX* genes have been shown to be transcriptionally regulated by *HOTAIR* [10, 35] in fibroblasts and breast cancer cell lines, it is currently unknown whether a similar regulation of *HOXA9* may also occur in glioma. Indeed, the expression of *HOXA9* in GBM is known to be epigenetically regulated by the PI3K pathway via inhibition of EZH2-mediated H3K27 trimethylation at the *HOXA9* locus [36], and *HOTAIR* is known to interact with PRC2 complex (containing EZH2 and other subunits) to guide it to target loci [10, 13, 37]. Therefore, it is conceptually feasible that *HOTAIR* might regulate *HOXA9* expression, which together with other transcription factors can regulate the expression of *HOTAIR*, in a putative loop mechanism. Of note, the roles of HOXA9 in glioma and in breast cancer are very distinct: while in the first it displays a variety of oncogenic functions, associated with tumor aggressiveness, higher grades of malignancy, tumor implantation in *in vivo* models, therapy resistance and shorter survival of patients [26, 38–40], in breast cancer HOXA9 is a tumor suppressor gene whose expression is frequently lost, leading to cell growth, survival, invasiveness and changes in morphogenesis [41]. These studies highlight the importance of cell context for the biological effects mediated by HOXA9, which do not thus preclude the possibility that the transcriptional regulatory mechanisms between *HOTAIR* and *HOXA9* may also be distinct in glioma and breast cancer. Although conceptually possible, this hypothesis has not been tested in GBM, and warrants further elucidation in new studies. We report here for the first time the ability of HOXA9, a homeoprotein, to directly regulate the expression of *HOTAIR*. While we provide strong indications that *HOTAIR*-positive GBMs may result from altered intragenic DNA methylation levels and/or HOXA9 binding, other mechanisms cannot be excluded. For example, the CT variant of the functional single nucleotide polymorphism rs12826786 in *HOTAIR* locus was associated with its higher intratumoral expression in glioma [42], and would require further addressing. Thus, our work provides novel insights on *HOTAIR*'s transcriptional regulators, widening our understanding of the *HOTAIR*-associated mechanisms of aggressiveness and malignancy in gliomas. Interpreting our findings at the light of other studies further supports the classic tissue-specific roles and regulatory mechanisms of lncRNAs [43].

The prognostic value of *HOTAIR* that we observed in 3 independent datasets further supports the reported clinical value of *HOTAIR* in glioma [16]. Critically, this is the first study establishing a prognostic value for *HOTAIR* in lower grade glioma (grades II and III). Importantly, we also found that *HOTAIR* expression was able to identify a subset of GBM patients with poor prognosis even in the universe of *HOXA9*-negative GBM patients, indicating that the combinatorial evaluation of *HOTAIR* and *HOXA9* expression levels may prove advantageous towards a molecularly-based stratification of GBM patients. Testing how the concomitant inhibition of *HOTAIR* and HOXA9 may affect the behavior of GBM *in vitro* and *in vivo* will be interesting and warrants further investigation in future studies. Furthermore, the clinical importance of *HOTAIR* suggests it as a putative therapeutic target. While drugs directly interfering with *HOTAIR* function are currently being studied [44–47] but still not clinically used in glioma, it may be interesting to test pharmacological approaches that act upstream of *HOTAIR* activation, namely PI3K inhibitors (e.g., PI-103 [48] and LY294002 [49]) that have been shown to silence *HOXA9* transcription, and might thus result in *HOTAIR* downregulation. In addition, part of *HOTAIR* effects in glioma may also be limited by using drugs directed to PRC2 components, as the EZH2 inhibitor DZNep [50], or drugs that specifically target LSD1 activity and have been tested in cancer [51, 52].

In summary, our study shows that (i) *HOTAIR* is highly expressed in a subset of malignant gliomas, independently of gene copy number alterations; (ii) epigenetic marks at the level of DNA methylation in particular CpG sites associate with *HOTAIR* levels in GBM; (iii) co-expression of *HOTAIR* and *HOXA9* occurs frequently in high-grade glioma, but not in other cancer types; (iv) *HOTAIR* is prognostically valuable in malignant glioma patients; and (v) *HOTAIR* and *HOXA9* may be useful biomarkers to integrate a molecularly-based stratification of GBM patients. We anticipate that *HOTAIR* may mediate some of the effector mechanisms by which HOXA9 creates a more aggressive and therapy-resistant form of GBM. Further studies are warranted to better identify *HOTAIR* downstream target genes at the genome-wide level in GBM, in an attempt to better understand the mechanisms by which *HOTAIR* affects survival of patients, and ultimately investigate new therapeutic opportunities.

## MATERIALS AND METHODS

### Glioma primary samples and cell lines

Glioma tumor specimens were obtained from "Hospital Braga" (*n* = 34) and "Hospitais da Universidade de Coimbra" (*n* = 39), Portugal, and from the Clermont-Ferrand University Hospital Center (*n* = 51), France (Supplementary Table 1), upon approval by each Hospital's ethical committee and patients signed a written individual informed consent according to institutional guidelines. Tumor tissues were snap-frozen in dry ice or isopentane cooled in liquid nitrogen directly from the operating room, and stored at –80° C. In the French dataset, 57% of the samples were IDH1 genotyped by EpigenDx (Worcester, MA) using pyrosequencing (assays ADS1703 and ADS1704), while the other 43% by analyzing the most frequent IDH1 mutations in exon 4 (codon 132) using RFLP [53]. Regarding 1p19q samples characterization, the analysis was performed using the Vysis 1p36/1q25 and 19q13/19p13 FISH probe kit (4N6020, Abbott), according to manufacturer's instructions. For some cases lacking 1p19q characterization, internexin A (INA) immunohistochemical analyses were carried out on paraffin sections using antibodies directed against internexin neuronal intermediate filament protein alpha (INA). Since INA expression was reported to be a surrogate marker of 1p19q codeletion [54, 55]), its expression was assessed to indirectly inquire about the 1p19q deletion status. Pediatric glioma cell lines KNS42, Res259, Res186, SF188, and UW479 were cultured in DMEM/F12 (Gibco, Grand Island, NY, USA) supplemented with 10% Fetal Bovine Serum (FBS; Biochrom, Berlin, DE), and 1% Penicillin-Streptomycin (Pen-Strep; Invitrogen, Grand Island, NY, USA) as previously described [56]. Commercially-available adult glioma cell lines A172, GAMG, SNB19, SW1783, U251, and U87MG were originally purchased from ATCC (Rockville, MD, USA), and the primary GBM-derived culture GBM-L18 (established in our lab) were cultured in DMEM (Gibco, Grand Island, NY, USA) supplemented with 10% FBS, and 1% Pen-Strep. U87MG (which do not express *HOXA9* endogenously) were previously [28] genetically-engineered with murine stem cell virus (MSCV) retroviral vectors containing the *HOXA9* coding region to obtain *HOXA9* overexpressing cells (U87MG-HOXA9) or with control empty vector (U87MG-MSCV). Selection of retrovirally-infected cells was maintained with 500 μg/mL G418 (Sigma-Aldrich, St. Louis, MO, USA). Incubations were performed at 37° C, in a humidified atmosphere containing 5% $CO_2$.

### TCGA data meta-analysis in glioma patients

*HOTAIR* gene expression, copy number, DNA methylation and clinical data were downloaded from TCGA; available from: http://cancergenome.nih.gov/, accessed 2017) [57].

TCGA Agilent's G4502A 244K gene expression profiles from 572 GBMs, 27 grades II and III gliomas (Supplementary Table 1), and 10 unmatched normal samples were analyzed, and "level 3" values of *HOTAIR* (probe A_32_P168442) and *HOXA9* (probe A_23_P500998) were used. *HOTAIR*-high was considered when "level 3" value ≥0, and *HOXA9*-high when ≥2. RNA-

www.oncotarget.com                                    15750                                    Oncotarget

seq data from additional 161 GBMs, 511 grades II and III glioma patients and 5 unmatched normal samples were also collected (Supplementary Table 1). These data were evaluated using Illumina HiSeq 2000 sequencing system. Genes with a "level 3" value = 0 were considered to have undetectable levels of mRNA expression. Cases were classified based on WHO malignancy grade and considering genetic information according to latest WHO recommendations [1] to: IDH-wt; IDH-mut (without 1p/19q codeletion), and IDH-mut and 1p/19q codeleted.

Gene copy number data were assessed by Affymetrix Genome-Wide Human SNP Array 6.0 in 372 GBM samples. Gene amplifications or deletions were considered for $Log_2$ copy number tumor/normal ≥0.5 (gene copy number ≥3) or ≤−0.5 (gene copy number ≤1), respectively.

Tumor DNA methylation profiles were detected by Illumina Infinium Human DNA Methylation 450 array and include the methylation status of 74 GBM samples. We evaluated 56 probes that span from the *HOXC12* (upstream of *HOTAIR*) to *HOXC11* (downstream of *HOTAIR*). High methylation was considered for β-values > 0.5.

### Analysis of *HOTAIR* expression in Oncomine and REMBRANDT

The human cancer microarray database Oncomine (*www.oncomine.com;* Oncomine™, Compendia Bioscience, Ann Arbor, MI) was used for analysis and visualization [58] of *HOTAIR* expression (probe 239153_at) in 45 WHO grade II, 98 WHO grade III, and 296 primary WHO grade IV gliomas (Supplementary Table 1). Categorization of *HOTAIR*-positive and *HOTAIR*-negative glioma patients was based on the $Log_2$ median-centered intensity values of *HOTAIR* probe. $Log_2$ median-centered intensity values >0 correspond to high *HOTAIR* expression, and $Log_2$ values ≤0 correspond to low/negative *HOTAIR* expression. The Oncomine database was also used to identify genes frequently co-expressed with *HOTAIR* in gliomas and other cancer datasets [lung ($n$ = 167), breast ($n$ = 593), leukemia ($n$ = 197), and colorectal cancer ($n$ = 237)]. Details on the normalization techniques and statistical calculations can be found on the website (*www.oncomine.com*) [58].

The REMBRANDT platform (National Cancer Institute REMBRANDT, accessed 2013 June; available now at https://gdoc.georgetown.edu/gdoc/) was queried to evaluate 67 GBM patients (Supplementary Table 1); the cut-off for *HOTAIR* high expression was established as ≥4-fold than *HOTAIR* expression in non-tumor samples.

### RNA extraction, cDNA synthesis, and gene expression analyses in primary glioma samples and cell lines

Total RNA was extracted from glioma patient samples and cell lines using the TRIzol method (Invitrogen, Carlsbad, CA, USA) for the Portuguese dataset or the RNeasy Mini Kit (Qiagen, Hamburg, GmbH) for the French dataset, and cDNA was synthesized using High Capacity cDNA Reverse Transcription kit (Applied Biosystems, Foster City, CA, USA) according to the manufacturer's protocol.

For the Portuguese dataset, gene expression levels were assessed by RT-PCR (AmpliTaq Gold 360, Applied Biosystems, Foster City, CA, USA) and quantitative reverse transcriptase real-time PCR (qPCR; KAPA Biosystems, Foster City, CA, USA) according to manufacturer's instructions (primer sequences in Supplementary Table 7). Housekeeping genes used to normalize gene expression levels were *hGUS*, *TBP*, *HPRT1*, or *ACTIN*, as stated on each analysis. For all tested genes, the standard PCR parameters were as follows: 4 minutes at 94° C, 35 cycles of denaturation for 30 seconds for the RT-PCR or 3 seconds for the qPCR at 94° C, annealing for 30 seconds (Supplementary Table 7), extension at 72° C for 30 seconds, and final extension at 72° C for 8 minutes for conventional PCR, or increments of 1° C each 5 seconds from 65 to 95° C for the qPCR. The cut-off for high expression of both *HOTAIR* and *HOXA9* was established as >5% of relative expression (upon normalization to housekeeping gene).

In the French dataset, microfluidic-based quantitative analysis (Fuidigm) was used to assess *HOTAIR* and *HOXA9* genes expression levels. cDNA was preamplified for 14 cycles with the pool of primers available in Supplementary Table 7 by using the TaqMan PreAmplification Master Mix (Applied Biosystems). qRT-PCRs were next conducted and validated on Fluidigm 96.96 Dynamic Arrays using the Biomark HD system (Fluidigm Corp.) according to the manufacturer's instructions. The relative quantification in gene expression was determined after normalization to a geometrical mean of housekeeping genes expression, namely *PPIA, HPRT1* and *TBP*. For each condition, data presented in this study are obtained from two independent experiments, each analyzed in duplicate.

### 5-Aza-2′-deoxycytidine treatment, DNA isolation, bisulfite conversion and *HOTAIR* methylation-specific PCR

Glioma cell lines were treated with 5 μM 5-Aza (Sigma-Aldrich®, St. Louis, MO, USA) or dimethyl sulfoxide (DMSO; Sigma-Aldrich®, St. Louis, MO, USA) for 72 hours with daily renewal. Cells were collected by trypsinization, and DNA and RNA were extracted by the TRIzol method. RNA was used to assess *HOTAIR* expression levels as previously described, and DNA was subjected to sodium bisulfite conversion using the EZ DNA Methylation-Gold™ Kit (Zymo Research, Foster City, CA, USA), according to the manufacturer's protocol. Touchdown MSPs for *HOTAIR* promoter were performed (AmpliTaq Gold 360; 3 cycles at 62° C with decrement of 1° C; 6 cycles at 59.5° C with decrement of 0.5° C; and 29 cycles at 57° C; primers in Supplementary Table 7). Blood DNA of a control subject (NB569) was bisulfite treated and used as an unmethylated control for MSP reactions. The same DNA was *in vitro* methylated (CpG Methyltransferase M.SssI; New England Biolabs Inc.) according to manufacturer's protocol, followed by sodium bisulfite treatment, and used as a methylated control.

In the French dataset, *HOTAIR* DNA methylation profile was obtained by Illumina Infinium Human DNA Methylation 450 array performed on the 43 glioma samples. We evaluated 26 probes that cover the *HOTAIR* locus between *HOXC11* and *HOXC12* (chr12:54,355,686–54,369,516). High methylation was considered for β-values > 0.5.

### *In silico* analysis of transcription factor binding sites in *HOTAIR* promoter by Genomatix

MatInspector from Genomatix software [59] (www.genomatix.de) was used to investigate putative transcription factors binding sites in *HOTAIR* gene. This *in silico* tool identifies transcription factor binding sites in nucleotide sequences based in a large library of weight matrices [59, 60]. A perfect match gets a matrix similarity of 1 when the tested sequence corresponds to the most conserved nucleotide at each position of the matrix. A good match to the matrix was considered when matrix similarity >0.80.

### Chromatin-immunoprecipitation analysis

ChIP experiments were done as previously described [61]. In brief, U251, U87MG-MSCV and U87MG-HOXA9 cells were cross-linked with 1.42% formaldehyde for 15 minutes, followed by quenching with 125 mM glycine for 5 minutes. Cells were then lysed with immunoprecipitation buffer (150 mM NaCl, 50 mM Tris-HCl, 5 mM EDTA, 0.5% NP-40, 1% Triton X-100) and chromatin was sheared by sonication (Sonics Vibra Cell VC70T, 21 cycles for 15 seconds) to obtain DNA fragments of approximately 0.5–1kb. The volume of sheared chromatin equivalent to 2 million cells was incubated with the required antibody in an ultrasonic bath for 15 minutes, followed by incubation with protein A-sepharose beads (Amersham, Uppsala, SE) and Chelex 100 (Bio-Rad, Hercules, CA, USA). The following antibodies were used per immunoprecipitation: 4 μg anti-HOXA9 (Santa Cruz Biotechnologies, Santa Cruz, CA), and 3 μg anti-Immunoglobulin G (IgG; Sigma-Aldrich, St. Louis, MO, USA) as ChIP negative control. The input represents a control for the amount of DNA used in precipitations. DNA amplification was done by qPCR (KAPA SYBR FAST, KAPA Biosystems) according to manufacturer's instructions with primers designed to amplify a portion of the *HOTAIR* promoter region that spans from bp -991 to -826 from the *HOTAIR* transcriptional start site (Supplementary Table 7). The qPCR parameters were as follows: 4 minutes at 94° C, 35 cycles of denaturation for 3 seconds, annealing for 30 seconds at 60° C, extension at 72° C for 30 seconds, and final extension consisting in increments of 1° C each 5 seconds from 65 to 95° C. The U251 fold enrichment of *HOTAIR* promoter over IgG, and the fold change occupancy of *HOTAIR* promoter in *HOXA9*-overexpressing over their negative counterparts for U87MG, were calculated for each experiment using the $\Delta\Delta C_t$ method as described previously [62]. Three biological replicates were tested, and each qPCR experiment was performed in triplicates.

### Statistical analyses

Statistical analyses were performed using the SPSS 22.0 software (SPSS, Inc., Chicago, IL, USA) for the TCGA and Portuguese datasets, and the Stata software, version 13 (StataCorp, College Station, TX, US) for the French dataset. Tests were two-sided, with a type I error set at α = 0.05. Due to more limited sample size in the French dataset, *HOTAIR* threshold has been determined according to ROC curve analysis, and estimation of several indexes recommended in literature [63–65] and biological relevance. Univariable survival analyses to assess the prognostic value of *HOTAIR* and of other clinicopathological features (patient age, gender, KPS, use of chemoradiation therapy, and institution where the patients were treated) were performed by the Log-rank test whenever these data were available. Additionally, the independent prognostic value of *HOTAIR* was further analyzed by a multivariable Cox proportional hazard model adjusted for those potential confounding variables. The Chi-square test was used to assess differences between the distributions of tumors with high and low *HOTAIR* expression, stratified for methylation levels (β-value, TCGA). Correlation between the expression levels of *HOTAIR* and *HOXA9* were calculated by Pearson's or Spearman's correlation as indicated throughout the text, and according to data normality. *HOTAIR* expression level differences between histological groups was assessed by one-way ANOVA (Kruskal-Wallis test) for the TCGA dataset. The qPCR differences in ChIP experiments and between histological groups in the French dataset, and genes expression were calculated by Mann-Whitney test and unpaired Student's *t*-test, respectively, using Prism GraphPad software (version 6.0a, San Diego, CA, USA). In Oncomine [58], each gene was evaluated for differential expression using Student's t-test in the case of two-class analyses (e.g. tumor tissue versus respective normal tissue); for multiclass analyses (e.g. grade II, III, and IV gliomas) Pearson's correlation was used. The association between methylation indexes of *HOTAIR* and its expression levels in the TCGA dataset was measured by the Pearson's correlation (r) using SPSS 22.0 software.

## Abbreviations

5-Aza: 5-Aza-2′-deoxycytidine; ChIP: Chromatin-immunoprecipitation; FBS: fetal bovine Serum; GBM: glioblastoma; HR: Hazard Ratio; H3K27: histone H3 lysine 27; H3K4me2: histone H3 dimethyl lysine 4; HOTAIR: HOX transcript antisense intergenic RNA; IDH: isocitrate dehydrogenase; IgG: Immunoglobulin G; lncRNA: long non-coding RNAs; KPS: Karnofsky Performance Score; LSD1/CoREST/REST: lysine specific demethylase 1/REST corepressor 1/RE1-silencing transcription factor; MSCV: murine stem cell virus; MSP: methylation-specific PCR; mut: mutant; OS: overall survival; Pen-Strep: penicillin-streptomycin; PRC2: polycomb repressor complex 2; qPCR: quantitative PCR; REMBRANDT: Repository of Molecular Brain Neoplasia Data; RT-PCR: reverse transcription-PCR; TCGA: The Cancer Genome Atlas; WHO: World Health Organization; wt: wildtype.

## Author contributions

A.X-M. and C.S.G. designed, conducted and analyzed experiments, and wrote the manuscript. A.F. and B.P. conducted experiments and analyzed data from the French dataset. A.F., T.L. and P.A. assisted in manuscript writing. T.L. and M.P. assisted in obtaining the glioma specimens and generation of DNA and RNA, and provided intellectual input. M.C.L., I.C., O.R. and H.T. provided the samples from the Hospital Coimbra dataset. J.L., R.M. and A.A.P. provided the samples from the Hospital Braga dataset. M.R., C.J., R.M.R, J.F.C., P.A., and N.S. assisted during data analysis and provided intellectual input and valuable discussion. B.M.C. designed and supervised experiments, analyzed data, and wrote the manuscript.

## ACKNOWLEDGMENTS

The authors would like to extend their appreciation to Sandro Queirós for helpful assistance regarding data processing from the TCGA.

## CONFLICTS OF INTEREST

The authors disclose no potential conflicts of interest.

## FUNDING

Fundação Para A Ciência e Tecnologia (PTDC/SAU-GMG/113795/2009; SFRH/BPD/33612/2009 and IF/00601/2012 to B.M.C.; SFRH/BD/88220/2012 to A.X.M.; SFRH/BD/92786/2013 to C.S.G; SFRH/BD/81042/2011 to M.P.; and SFRH/BD/51996/2012 to T.L.), Project co-financed by Programa Operacional Regional do Norte (ON.2 – O Novo Norte), Quadro de Referência Estratégico Nacional (QREN), Fundo Europeu de Desenvolvimento Regional (FEDER); Fundação Calouste Gulbenkian (B.M.C.); and Liga Portuguesa Contra o Cancro, Portugal (B.M.C.). This article has been developed under the scope of the projects NORTE-01-0246-FEDER-000012, NORTE-01-0145-FEDER-000023 and NORTE-01-0145-FEDER-000013, supported by the Northern Portugal Regional Operational Programme (NORTE 2020), under the Portugal 2020 Partnership Agreement, through the European Regional Development Fund (FEDER). This work has been funded by FEDER funds, through the Competitiveness Factors Operational Programme (COMPETE), and by National funds, through the Foundation for Science and Technology (FCT), under the scope of the project POCI-01-0145-FEDER-007038. C.J. acknowledges NHS funding to the Biomedical Research Centre. P.A. acknowledges the Plan Cancer-INSERM (CS14085CS'Gliobiv', PA), the Cancéropole CLARA (Oncostarter «Gliohoxas»; PA), Fonds de dotation Patrick Brou de Lauriére (PA).

29. Gaspar N, Marshall L, Perryman L, Bax DA, Little SE, Viana-Pereira M, Sharp SY, Vassal G, Pearson AD, Reis RM, Hargrave D, Workman P, Jones C. MGMT-independent temozolomide resistance in pediatric glioblastoma cells associated with a PI3-kinase-mediated HOX/stem cell gene signature. Cancer research. 2010; 70:9243–52. https://doi.org/10.1158/0008-5472.CAN-10-1250.

30. Gupta RA, Shah N, Wang KC, Kim J, Horlings HM, Wong DJ, Tsai MC, Hung T, Argani P, Rinn JL, Wang Y, Brzoska P, Kong B, et al. Long non-coding RNA HOTAIR reprograms chromatin state to promote cancer metastasis. Nature. 2010; 464:1071–6. https://doi.org/nature0897510.1038/nature08975.

31. Zhou X, Ren Y, Zhang J, Zhang C, Zhang K, Han L, Kong L, Wei J, Chen L, Yang J, Wang Q, Zhang J, Yang Y, et al. HOTAIR is a therapeutic target in glioblastoma. Oncotarget. 2015; 6:8353–65. https://doi.org/10.18632/oncotarget.3229.

32. Bian EB, Ma CC, He XJ, Wang C, Zong G, Wang HL, Zhao B. Epigenetic modification of miR-141 regulates SKA2 by an endogenous 'sponge' HOTAIR in glioma. Oncotarget. 2016; 7:30610–25. https://doi.org/10.18632/oncotarget.8895.

33. Maunakea AK, Nagarajan RP, Bilenky M, Ballinger TJ, D'Souza C, Fouse SD, Johnson BE, Hong C, Nielsen C, Zhao Y, Turecki G, Delaney A, Varhol R, et al. Conserved role of intragenic DNA methylation in regulating alternative promoters. Nature. 2010; 466:253–7. https://doi.org/nature0916510.1038/nature09165.

34. Lorincz MC, Dickerson DR, Schmitt M, Groudine M. Intragenic DNA methylation alters chromatin structure and elongation efficiency in mammalian cells. Nature structural & molecular biology. 2004; 11:1068–75. https://doi.org/10.1038/nsmb840.

35. Wang L, Zeng X, Chen S, Ding L, Zhong J, Zhao JC, Sarver A, Koller A, Zhi J, Ma Y, Yu J, Chen J, Huang H. BRCA1 is a negative modulator of the PRC2 complex. The EMBO journal. 2013; 32:1584–97. https://doi.org/10.1038/emboj.2013.95.

36. Cancer Genome Atlas Research Network. Comprehensive genomic characterization defines human glioblastoma genes and core pathways. Nature. 2008; 455:1061–8. https://doi.org/10.1038/nature07385.

37. Khalil AM, Guttman M, Huarte M, Garber M, Raj A, Rivea Morales D, Thomas K, Presser A, Bernstein BE, van Oudenaarden A, Regev A, Lander ES, Rinn JL. Many human large intergenic noncoding RNAs associate with chromatin-modifying complexes and affect gene expression. Proc Natl Acad Sci USA. 2009; 106:11667–72. https://doi.org/10.1073/pnas.0904715106.

38. Costa BM, Smith JS, Chen Y, Chen J, Phillips HS, Aldape KD, Zardo G, Nigro J, James CD, Fridlyand J, Reis RM, Costello JF. Reversing HOXA9 oncogene activation by PI3K inhibition: epigenetic mechanism and prognostic significance in human glioblastoma. Cancer Res. 2010; 70:453–62. https://doi.org/0008-5472.CAN-09-218910.1158/0008-5472.CAN-09-2189.

39. Murat A, Migliavacca E, Gorlia T, Lambiv WL, Shay T, Hamou MF, de Tribolet N, Regli L, Wick W, Kouwenhoven MC, Hainfellner JA, Heppner FL, Dietrich PY, et al. Stem cell-related "self-renewal" signature and high epidermal growth factor receptor expression associated with resistance to concomitant chemoradiotherapy in glioblastoma. J Clin Oncol. 2008; 26:3015–24. https://doi.org/26/18/301510.1200/JCO.2007.15.7164.

40. Gaspar N, Marshall L, Perryman L, Bax DA, Little SE, Viana-Pereira M, Sharp SY, Vassal G, Pearson AD, Reis RM, Hargrave D, Workman P, Jones C. MGMT-independent temozolomide resistance in pediatric glioblastoma cells associated with a PI3-kinase-mediated HOX/stem cell gene signature. Cancer Res. 2010; 70:9243–52. https://doi.org/0008-5472.CAN-10-125010.1158/0008-5472.CAN-10-1250.

41. Gilbert PM, Mouw JK, Unger MA, Lakins JN, Gbegnon MK, Clemmer VB, Benezra M, Licht JD, Boudreau NJ, Tsai KK, Welm AL, Feldman MD, Weber BL, et al. HOXA9 regulates BRCA1 expression to modulate human breast tumor phenotype. J Clin Invest. 2010; 120:1535–50. https://doi.org/10.1172/JCI39534.

42. Xavier-Magalhaes A, Oliveira AI, de Castro JV, Pojo M, Goncalves CS, Lourenco T, Viana-Pereira M, Costa S, Linhares P, Vaz R, Nabico R, Amorim J, Pinto AA, et al. Effects of the functional HOTAIR rs920778 and rs12826786 genetic variants in glioma susceptibility and patient prognosis. J Neurooncol. 2017; 132:27–34. https://doi.org/10.1007/s11060-016-2345-0.

43. Mercer TR, Dinger ME, Sunkin SM, Mehler MF, Mattick JS. Specific expression of long noncoding RNAs in the mouse brain. Proc Natl Acad Sci U S A. 2008; 105:716–21. https://doi.org/10.1073/pnas.0706729105.

44. Pastori C, Kapranov P, Penas C, Peschansky V, Volmar CH, Sarkaria JN, Bregy A, Komotar R, St Laurent G, Ayad NG, Wahlestedt C. The Bromodomain protein BRD4 controls HOTAIR, a long noncoding RNA essential for glioblastoma proliferation. Proc Natl Acad Sci U S A. 2015; 112:8326–31. https://doi.org/10.1073/pnas.1424220112.

45. Jiang Y, Zhang Q, Bao J, Du C, Wang J, Tong Q, Liu C. Schisandrin B inhibits the proliferation and invasion of glioma cells by regulating the HOTAIR-micoRNA-125a-mTOR pathway. Neuroreport. 2017; 28:93–100. https://doi.org/10.1097/WNR.0000000000000717.

46. Chen J, Lin C, Yong W, Ye Y, Huang Z. Calycosin and genistein induce apoptosis by inactivation of HOTAIR/p-Akt signaling pathway in human breast cancer MCF-7 cells. Cell Physiol Biochem. 2015; 35:722–8. https://doi.org/10.1159/000369732.

47. Chiyomaru T, Yamamura S, Fukuhara S, Yoshino H, Kinoshita T, Majid S, Saini S, Chang I, Tanaka Y, Enokida H. Genistein inhibits prostate cancer cell growth by targeting miR-34a and oncogenic HOTAIR. PloS One. 2013; 8:e70372.
www.oncotarget.com  15755  Oncotarget